\documentclass[12pt]{article}
\usepackage[english]{babel}
\usepackage{graphicx}

\begin{document}
\textsl{}
\begin{center}
\begin{large}
\textbf{STRUCTURE AND DIFFERENT REALIZATIONS OF THE EXTENDED REAL CLIFFORD--DIRAC ALGEBRA}
\end{large}
\vskip 0.5cm

\textbf{V.M. Simulik$^1$, I.Yu. Krivsky$^1$, I.O. Gordievich$^2$, I.L. Lamer$^1$}
\end{center}

\begin{center}
\textit{$^1$Institute of Electron Physics, National Academy of
Sciences of Ukraine, 21 Universitetska Str., 88000 Uzhgorod,
Ukraine}

\textit{$^2$Technocristal Corund, Uzhgorod, Ukraine}

\end{center}

\begin{center}
\textit{E-mail: vsimulik@gmail.com}
\end{center}

\vskip 1.cm

\noindent ABSTRACT. The structure of the 64-dimensional extended real Clifford--Dirac (ERCD) algebra, which has been introduced in our paper Phys. Lett. A. 375 (2011) 2479, is under consideration. The subalgebras of this algebra are investigated: the 29-dimensional proper ERCD algebra and 32-dimensional pure matrix algebra of invariance of the Dirac equation in the Foldy--Wouthuysen representation. The last one is the maximal pure matrix algebra of invariance of this equation. The different realizations of the proper ERCD algebra are given. The application of proper ERCD algebra is illustrated on the example of the derivation of the hidden spin (1,0) Poincare symmetry of the Dirac equation.

\vskip 0.5cm


\vskip 0.5cm


\vskip 1.cm

\section{Introduction}

This paper has been prepared in June 2012 for the Proceedings of the BGL-8 International Conference on Non-Euclidean Geometry in Modern Physics and Mathematics. The conference was held in Ukraine, Uzhgorod, 22--25 May 2012 in the Institute of Electron Physics. The fast publication of the Proceedings was promised in October 2012 in the next coming special issue of the Uzhgorod University Scientific Herald, Ser. Physics. Today it is evident that this Proceedings never will be published and the promises of the Organizing committee are groundless.

That is the reason of our presentation of this paper here in the form of e-preprint.  Better later than never. 

The paper below is an extended version of brief article [1], which was reported at the BGL-8 Internationalional Conference on Non-Euclidean Geometry in Modern Physics and Mathematics and published in the
Book of abstracts of this conference.

The system of units $\hbar=c=1$ and metric $g=(g^{\mu\nu})=(+---), \, a^{\mu}=g^{\mu\nu}a_{\nu},$ are taken. Here the Greek indices are changed in the region $0,1,2,3=\overline{0,3}$, Latin -- $\overline{1,3}$, the summation over the twice repeated index is implied.

Our consideration is fulfilled in the
rigged Hilbert space
$\mathrm{S}^{3,4}\subset\mathrm{H}^{3,4}\subset\mathrm{S}^{3,4*}$
where $\mathrm{H}^{3,4}$ is given by

\begin{equation}
\label{eq1}
\mathrm{H}^{3,4}=\mathrm{L}_{2}(\mathrm{R}^3)\otimes\mathrm{C}^{\otimes4}=\{\phi=(\phi^{\mu}):\mathrm{R}^{3}\rightarrow\mathrm{C}^{\otimes4};
 \, \int
d^{3}x|\phi(t,\overrightarrow{x})|^{2} <\infty\} \},
\end{equation}

\noindent and symbol "*" in $\mathrm{S}^{3,4*}$ means, that the
space of Schwartz generalized functions $\mathrm{S}^{3,4*}$ is
conjugated to the Schwartz test function space $\mathrm{S}^{3,4}$
by the corresponding topology. The application of the rigged Hilbert space allows one to reproduce a detailed consideration of a field theory in mathematically correct form.
In order to apply our generalization [2] let us start from the brief consideration of the standard Clifford--Dirac (CD) algebra. The Dirac matrices

\begin{equation}
\label{eq2}
\gamma^{\mu}\, (\mu=\overline{0,3}):\gamma^{\mu}\gamma^{\nu}+\gamma^{\nu}\gamma^{\mu}=2g^{\mu\nu},\, \gamma_{\mu}=g_{\mu\nu}\gamma^{\nu}, \, \gamma^{\dag k}=-\gamma^{k}, \, \gamma^{\dag 0}=\gamma^{0},
\end{equation}

\noindent we chose in the standard Pauli--Dirac (PD) representation

\begin{equation}
\label{eq3}
\gamma ^0 = \left| {{\begin{array}{*{20}c}
 1 \hfill & 0 \hfill \\
 0 \hfill & { - 1} \hfill \\
\end{array} }} \right|, \, \gamma ^k = \left| {{\begin{array}{*{20}c}
 0 \hfill & {\sigma ^k} \hfill \\
 { - \sigma ^k} \hfill & 0 \hfill \\
\end{array} }}\right|, \quad k = \overline{1,3},
\end{equation}

\noindent where the Pauli matrices are given by

\begin{equation}
\label{eq4}
\sigma ^1 = \left| {{\begin{array}{*{20}c}
 0 \hfill & 1 \hfill \\
 1 \hfill & 0 \hfill \\
\end{array} }} \right|, \, \sigma ^2 = \left| {{\begin{array}{*{20}c}
 0 \hfill & {-i} \hfill \\
 i \hfill & 0 \hfill \\
\end{array} }}\right|, \, \sigma ^3 = \left| {{\begin{array}{*{20}c}
 1 \hfill & 0 \hfill \\
 0 \hfill & -1 \hfill \\
\end{array} }}\right|.
\end{equation}

We redefine the matrix, which if often used in the literature as $\gamma^{5}$, and chose this matrix in more convenient form

\begin{equation}
\label{eq5}
\gamma ^4  = \gamma ^0 \gamma ^1 \gamma ^2 \gamma ^3 = -i \left| {{\begin{array}{*{20}c}
 0 \hfill & \mathrm{I} \hfill \\
 \mathrm{I} \hfill & 0 \hfill \\
\end{array} }} \right|, \quad \mathrm{I} = \left| {{\begin{array}{*{20}c}
 1 \hfill & 0 \hfill \\
 0 \hfill & 1 \hfill \\
\end{array} }}\right|. 
\end{equation}

\noindent After that for the 5 $\gamma$ matrices the following anticommutation relations

\begin{equation}
\label{eq6}
\gamma^{\bar{\mu}}\gamma^{\bar{\nu}}+\gamma^{\bar{\nu}}\gamma^{\bar{\mu}}=2g^{\bar{\mu}\bar{\nu}}; \quad \bar{\mu},\bar{\nu} = \overline{0,4}, \, (g^{\bar{\mu}\bar{\nu}})=(+----),
\end{equation}

\noindent of the CD algebra are hold.

\section{Motivations}

In general the relationship of the Clifford algebra to the algebras of SO(m,n) groups was investigated by E. Cartan. In our consideration of the CD algebra we refer on the papers [3, 4], where the following assertion was proved. \textit{The 16 orts of CD algebra realize the representation of the} SO(3,3) \textit{algebra}. For our purposes [2] it was useful to rewrite this assertion in the form of $\mathrm{SO(1,5)}\supset \mathrm{SO(1,3)}$ algebra, similarly to the algebra of rotations in the 6-dimensional Minkowski space $\mathrm{M(1,5)}\supset \mathrm{M(1,3)}$ (in analogy with the representations of the proper ortochronous Lorentz group $\mbox{L}_ + ^
\uparrow = \mathrm{SO(1,3)}$ ) in the space M(1,3)). Here M(1,5) is the extension of the real physical 4-dimensional space-time M(1,3) into 6-dimensional pseudo-Euclidean space-time.

Hence, it is useful to chose the 16 independent orts $\alpha^{\tilde{\mu}\tilde{\nu}}+ \mathrm{I}$ of the standard CD algebra, connected with the SO(1,5) algebra generators, in the form

\begin{equation}
\label{eq7}
\left\{ {\mbox{ind \, CD}} \right\} \equiv \{
\mbox{I}, \, \alpha^{\tilde{\mu}\tilde{\nu}} =
2{s}^{\tilde{\mu}\tilde{\nu}} \},  \quad \tilde{\mu},\tilde{\nu} = \overline{0,5},
\end{equation}

\noindent where

\begin{equation}
\label{eq8} {s}^{\bar{\mu}\bar{\nu}} \equiv \frac{1}{4}\left[
\gamma^{\bar{\mu}},\gamma^{\bar{\nu}} \right],\mbox{
}{s}^{\bar{\mu}5} = - {s}^{5\bar{\mu}}\equiv
\frac{1}{2}\gamma^{\bar{\mu}};\mbox{ }\gamma^{4}\equiv
\gamma^{0}\gamma^{1}\gamma^{2}\gamma^{3},\mbox{ }
\bar{\mu},\bar{\nu}=\overline{0,4}.
\end{equation}

\noindent The matrices (8) satisfy the commutation relations of SO(1,5) algebra in the following form

\begin{equation}
\label{eq9}\left[ {{s}^{\tilde{\mu}\tilde{\nu}},
{{s}^{\tilde{\rho}\tilde{\sigma}}}}\right] = -
{g}^{\tilde{\mu}\tilde{\rho}} {s}^{\tilde{\nu}\tilde{\sigma}} -
{g}^{\tilde{\rho}\tilde{\nu}} {s}^{\tilde{\sigma}\tilde{\mu}} -
{g}^{\tilde{\nu}\tilde{\sigma}} {s}^{\tilde{\mu}\tilde{\rho}} -
{g}^{\tilde{\sigma}\tilde{\mu}}
{s}^{\tilde{\rho}\tilde{\nu}};\mbox{ }
(g_{\tilde{\nu}}^{\tilde{\mu}})=\mathrm{diag}(+1,-1,-1,-1,-1,-1).
\end{equation}

Four operators $\gamma^{\mu} \, (\mu = \overline{0,3})$, which completely determine the set of 16 orts (7) of standard CD algebra, we call \textit{the prime generators} of this algebra. Furthermore, the 5 operators

\begin{equation}
\label{eq10} {s}^{\bar{\mu}5} = - {s}^{5\bar{\mu}}\equiv
\frac{1}{2}\gamma^{\bar{\mu}}, \mbox{ }
\bar{\mu},\bar{\nu}=\overline{0,4},
\end{equation}

\noindent due to their properties (10) we call as \textit{the generating orts} of SO(1,5) algebra.

We consider the CD algebra as the algebra over the field of real numbers due to the following reasons. We take into account that the proper ortochronous Lorentz $\mbox{L}_ + ^
\uparrow = \mathrm{SO(1,3)}$ and Poincar$\mathrm{\acute{e}}$ $\mbox{P}_ + ^
\uparrow$ groups are the real parametric groups Lie (we chose the tensor $\varpi = (\varpi_{\mu\nu})$ of rotations and the vector $a = (a_{\mu})$ of translations in the Minkowski space M(1,3) as the real parameters of the groups $\mbox{P}_ + ^
\uparrow \supset \mbox{L}_ + ^
\uparrow$). Therefore, the algebras of these groups are real too. Hence, we consider here the extensions of the SO(1,3) algebra as a real ones. Just due to the definitions (7), (8) and commutation relations (9) the CD algebra
$\mathrm{SO(1,5)}\supset \mathrm{SO(1,3)}$ is the real linear space spanned on the orts (7), (8)). Furthermore, in connection with real parameters, we consider the real anti-Hermitian generators of the corresponding groups and algebras (Lorentz, Poincar$\mathrm{\acute{e}}$ groups and CD algebra). For example, the energy-momentum $p^{\mu}$ and spin $s^{\ell n}$ operators in the Hilbert space (1) we take in the form

\begin{equation}
\label{eq11}
p^{\mu} = \partial ^{\mu}, \, s^{\ell n} = \frac{1}{4}[\gamma^{\ell},\gamma^{n}],
\end{equation}

\noindent (and not as $i\partial ^{\mu}, \, \frac{i}{4}[\gamma^{\ell},\gamma^{n}]$, respectively). Hence, the $\alpha^{\tilde{\mu}\tilde{\nu}}$ CD orts in (7), (8) are anti-Hermitian operators in the space (1) too. The application of such generators and mathematical correctness of corresponding formalism has been considered in details in [5, 6].

In addition to the CD algebra generators (7), (8) let us consider the set of following specific pure matrix operators

\begin{equation}
\label{eq12}
\left\{\gamma^{2}\hat{C},\,i\gamma^{2}\hat{C},\,\gamma^{2}\gamma^{4}\hat{C},\,i\gamma^{2}\gamma^{4}\hat{C},\,\gamma^{4},\,i\gamma^{4},
\,i,\,\mathrm{I}\right\}
\end{equation}

\noindent where $\hat{C}$ is the operator of complex conjugation, $\hat{C}\phi = \phi^{*}$, (i. e. $\hat{C}$ is the operator of involution in the space $\mathrm{H}^{3,4}$). Operators (12) generate the real algebra $\mathrm{A}_{8}$ of the Pauli--Gursey--Ibragimov [7]. It is known from [7] that the massless Dirac equation is invariant with respect to the pure matrix transformation from $\mathrm{A}_{8}$. 

In our papers, see e. g. [8--10] and the references therein, the following assertion has been proved. Part of the operators (12) realizes the additional $\mathrm{D}\left(0,\frac{1}{2}\right)\oplus \left(\frac{1}{2},0\right)$ representation of the universal covering algebra $\mathcal{L}=$SL(2,C) of proper ortochronous Lorentz group $\mathrm{L}_ + ^\uparrow =$SO(1,3). These 6 generators are given by

\begin{equation}
\label{eq13}
\left\{s^{01}_{\mathrm{PGI}}=\frac{i}{2}\gamma^{2}\hat{C},\,s^{02}_{\mathrm{PGI}}=-\frac{1}{2}\gamma^{2}\hat{C},\,s^{03}_{\mathrm{PGI}}=-\frac{i}{2}\gamma^{4},\,s^{23}_{\mathrm{PGI}}=\frac{i}{2}\gamma^{2}\gamma^{4}\hat{C},\,s^{31}_{\mathrm{PGI}}=-\frac{1}{2}\gamma^{2}\gamma^{4}\hat{C},\,s^{12}_{\mathrm{PGI}}=-\frac{i}{2}\right\}
\end{equation}

The simplest linear combinations of generators $s^{\mu\nu}$ from (8) and $s^{\mu\nu}_{\mathrm{PGI}}$ (13) enabled us [8] to construct the new bosonic representations of the Lorentz group $\mathcal{L}$ and Poincar$\mathrm{\acute{e}}$ group $\mathcal{P}\supset\mathcal{L}=$SL(2,C) (where $\mathcal{P}$ is the universal covering of the proper ortochronous Poincare group $\mathrm{P}_ + ^\uparrow \supset \mathrm{L}_ + ^\uparrow =$SO(1,3)), with respect to which the massless Dirac equation is invariant. The corresponding bosonic $\mathrm{D}(1,0)\oplus (0,0)$, $\mathrm{D}(\frac{1}{2},\frac{1}{2})$ representations of the algebra of the Lorentz group $\mathcal{L}$ together with tensor-scalar spin (1,0) and vector representations of the algebra of the Poincar$\mathrm{\acute{e}}$ group $\mathcal{P}$ have been found as the algebras of invariance of the massless Dirac equation. Moreover, similarly the slightly generalized original Maxwell equations have been proved [9] to be invariant with respect to fermionic spin $\frac{1}{2}$ representations of these groups. On this basis different relations between the solutions, conservation laws and other characteristics of the Dirac and Maxwell equations were found. These results of our group are well cited and explored by many authors, see e. g. [11--16] (the review of Prof. J. Keller [11] and close to practice using of Rozzi, Mencarelli and Pierantoni [16] are the most valuable for us).

The problem appeared \textit{"Is it a specific of zero mass or how to extend these results for the general case of nonzero mass in the Dirac and corresponding Maxwell equations?"}

In order to extend our consideration for the general case, when the mass in the Dirac equation is nonzero, we started to search for the complete set of operators of CD (7), (8) and Pauli--Gursey--Ibragimov type (12); and to search the corresponding generalization of CD algebra. This way brings us to our extension [2] of the CD algebra.

\section{Extended real Clifford--Dirac algebra}

Extended real Clifford--Dirac (ERCD) algebra has been found [2] as the complete set of operators of CD (7), (8) and Pauli--Gursey--Ibragimov type (12). The maximal extension of the real CD algebra can be constructed with the help of imaginary unit $i=\sqrt{-1}$ and the operation $\hat{C}$ of complex conjugation being considered as the operators in $\mathrm{H}^{3,4}$. In the terms of 16  orts (5) of standard CD algebra the ERCD algebra is written as

\begin{equation}
\label{eq14} \left\{ {\mbox{ERCD}} \right\} = \left\{ ({\mbox{ind
\, CD}}) \cup i\cdot({\mbox{ind \, CD}}) \cup
\hat{C}\cdot({\mbox{ind \, CD}}) \cup i\hat{C}\cdot({\mbox{ind \,
CD}}) \right\}.
\end{equation}

\noindent It is evident from the definitions (7), (8) and (14) that ERCD algebra contains 64 independent elements -- the orts of this algebra. Hence, the ERCD algebra as the real algebra in the space $\mathrm{H}^{3,4}$, includes the linear combinations with real parameters of all possible compositions of $\gamma^{\bar{\mu}}$ matrices, operators of imaginary unit $i=\sqrt{-1}$ and $\hat{C}$. Note that operators (14), orts of ERCD algebra, have the same properties as the orts of standard CD algebra. They commute or anticommute between each other and the square of each ort is equal to +I or -I. It is the reason of our name ERCD algebra for this mathematical structure.

Here six operators $\gamma^{\mu} \, (\mu =\overline{0,3}) $, imaginary unit $i=\sqrt{-1}$ and $\hat{C}$ are the prime generators of ERCD algebra and completely determine the set (14) of 64 orts of this algebra.

All the additional symmetry properties of the Dirac equation with nonzero mass, which have been found in [2], were constructed by using the elements of the ERCD algebra.

Among the 64 independent generators of the ERCD algebra one can find \textit{36 Hermitian and 28 anti-Hermitian orts.} Namely the set of last 28 operators generates the subalgebra SO(8) -- the proper ERCD algebra.

\section{Subalgebras of the extended real Clifford--Dirac algebra}

It is evident that the algebra $\mathrm{A}_{8}$ of the Pauli--Gursey--Ibragimov [7] is the subalgebra of ERCD algebra. The algebra SO(1.3) (13) is the subalgebra of ERCD algebra as well. However, the most important subalgebra of ERCD algebra is the 29 dimensional proper ERCD algebra [2].

\textbf{Proper extended Clifford--Dirac algebra} (PERCD) is the direct real generalization of standard CD algebra. The 29 orts of the PERCD algebra SO(8) open new possibilities in comparison with only 16 orts of SO(1.5) algebra, which form the standard CD algebra.

Now we have not 5, as it was in common Dirac theory, but 7 generating matrices (we also fixed them in the standard PD representation):

\begin{equation}
\label{eq15}
\left\{\gamma^{\mathrm{A}}\right\}=\left\{\gamma^{1},\,\gamma^{2},\,\gamma^{3},\,\gamma^{4}=\gamma^{0}\gamma^{1}\gamma^{2}\gamma^{3},\,\gamma^{5}=\gamma^{1}\gamma^{3}\hat{C},
\,\gamma^{6}=i\gamma^{1}\gamma^{3}\hat{C},\,\gamma^{7}=i\gamma^{0} \right\}.
\end{equation}

\noindent Matrices (15) obey the anticommutation relations in the form

\begin{equation}
\label{eq16} \gamma ^\mathrm{A} \gamma ^\mathrm{B} + \gamma
^\mathrm{B}\gamma ^\mathrm{A} = -
2\delta^{\mathrm{A}\mathrm{B}},\;\mathrm{A},\mathrm{B}=\overline{1,7},
\end{equation}

\noindent and generate the 28 orts $\alpha^{\bar{\mathrm {A}}\bar{\mathrm {B}}} =
2{s}^{\bar{\mathrm {A}}\bar{\mathrm {B}}}$:

\begin{equation}
\label{eq17}
s^{\bar{\mathrm{A}}\bar{\mathrm{B}}}=\{s^{\mathrm{A}\mathrm{B}}=\frac{1}{4}[\gamma
^\mathrm{A},\gamma
^\mathrm{B}],\,s^{\mathrm{A}8}=-s^{8\mathrm{A}}=\frac{1}{2}\gamma
^\mathrm{A}\},\,\bar{\mathrm{A}},\bar{\mathrm{B}}=\overline{1,8},
\end{equation}

\noindent which satisfy the commutation relations of SO(8) algebra

\begin{equation}
\label{eq18}
[s^{\bar{\mathrm{A}}\bar{\mathrm{B}}},s^{\bar{\mathrm{C}}\bar{\mathrm{D}}}]=
\delta^{\bar{\mathrm{A}}\bar{\mathrm{C}}}s^{\bar{\mathrm{B}}\bar{\mathrm{D}}}
+\delta^{\bar{\mathrm{C}}\bar{\mathrm{B}}}s^{\bar{\mathrm{D}}\bar{\mathrm{A}}}
+\delta^{\bar{\mathrm{B}}\bar{\mathrm{D}}}s^{\bar{\mathrm{A}}\bar{\mathrm{C}}}
+\delta^{\bar{\mathrm{D}}\bar{\mathrm{A}}}s^{\bar{\mathrm{C}}\bar{\mathrm{B}}}.
\end{equation}

As the consequences of the equalities

\begin{equation}
\label{eq19}
\gamma^{4}\equiv \prod^{3}_{\mu=0}\gamma^{\mu} \rightarrow \prod^{4}_{\bar{\mu}=0}\gamma^{\bar{\mu}} = -\mathrm{I},
\end{equation}

\noindent known from the standard CD algebra, and the anticommutation relations (16), in PERCD algebra for the matrices $\gamma^{\mathrm{A}}$ (15) the following extended equalities are valid:

\begin{equation}
\label{eq20}
\gamma^{7}\equiv -\prod^{6}_{\underline{\mathrm{A}}=1}\gamma^{\underline{\mathrm{A}}} \rightarrow \prod^{7}_{\mathrm{A}=1}\gamma^{\mathrm{A}} = \mathrm{I}, \quad \gamma^{5}\gamma^{6}=i.
\end{equation}

Note that subalgebra SO(6)$\subset$SO(8) of ERCD algebra is the algebra of invariance of the Dirac equation in the Foldy--Wouthuysen (FW) representation [17, 18]. The 16 orts of this SO(6) algebra are given by

\begin{equation}
\label{eq21}
\left\{\mathrm{I}, \, \alpha^{\underline{\mathrm{A}} \,\underline{\mathrm{B}}} =
2{s}^{\underline{\mathrm{A}} \, \underline{\mathrm{B}}}\right\},  \quad  \underline{\mathrm{A}},\underline{\mathrm{B}} = \overline{1,6},
\end{equation}

\noindent where

\begin{equation}
\label{eq22}
\left\{s^{\underline{\mathrm{A}} \,\underline{\mathrm{B}}}\right\}=\left\{s^{\underline{\mathrm{A}} \,\underline{\mathrm{B}}}\equiv \frac{1}{4}{\left[\gamma^{\underline{\mathrm{A}}},\gamma^{\underline{\mathrm{B}}}\right]}\right\}. 
\end{equation}

\noindent Now only the first 6 matrices

\begin{equation}
\label{eq23}
\left\{\gamma^{\underline{\mathrm{A}}}\right\}=\left\{\gamma^{1},\gamma^{2},\gamma^{3},\gamma^{4},\gamma^{5},\gamma^{6}\right\} 
\end{equation}

\noindent from the set (14) play the role of generating operators in the constructions (21), (22).

Note that the realization (21), (22) of SO(6) algebra is not unique. In other possible realizations instead of 5 matrices $\alpha^{1\underline{\mathrm{A}}}$ from (21) the matrices from $\left\{\gamma^{\underline{\mathrm{A}}}\right\}$ (23) can be used. However, in such realizations the operators from $\left\{\gamma^{\underline{\mathrm{A}}}\right\}$ (23) are not the transformations of invariance of the FW equation [17, 18].

The additional orts of algebra SO(6) (22), in comparison with standard CD algebra elements $\alpha^{mn}, \, \alpha^{m4}$, in the terms of standard $\gamma^{\mu}$ matrices have the form:

$$\alpha^{15}=-\gamma^{3}\hat{C}, \, \alpha^{25}=-\gamma^{0}\gamma^{4}\hat{C},\, \alpha^{35}=\gamma^{1}\hat{C}, \,\alpha^{45}=\gamma^{0}\gamma^{2}\hat{C},$$
\begin{equation}
\label{eq24}
\alpha^{16}=-i\gamma^{3}\hat{C}, \, \alpha^{26}=-i\gamma^{0}\gamma^{4}\hat{C},\, \alpha^{36}=i\gamma^{1}\hat{C}, \,\alpha^{46}=i\gamma^{0}\gamma^{2}\hat{C}, \, \alpha^{56}=i. 
\end{equation}

\noindent Other new elements of SO(8) (17) in the terms of $\gamma^{\mu}$ (3) are given by

$$\alpha^{17}=-i\gamma^{0}\gamma^{1}, \, \alpha^{27}=-i\gamma^{0}\gamma^{2},\, \alpha^{37}=-i\gamma^{0}\gamma^{3}, \,\alpha^{47}=-i\gamma^{0}\gamma^{4}, \, \alpha^{57}=-i\gamma^{2}\gamma^{4}\hat{C}, \, \alpha^{67}=\gamma^{2}\gamma^{4}\hat{C},$$
\begin{equation}
\label{eq25}
\alpha^{18}=\gamma^{1}, \,\alpha^{28}=\gamma^{2}, \, \alpha^{38}=\gamma^{3}, \, \alpha^{48}=\gamma^{4}, \, \alpha^{58}=\gamma^{1}\gamma^{3}\hat{C}, \,\alpha^{68}=i\gamma^{1}\gamma^{3}\hat{C}, \, \alpha^{78}=i\gamma^{0}. 
\end{equation}

\noindent Formulae (24) give the explicit forms of the orts both SO(6) and PERCD SO(8) algebras. Formulae (25) contain only the orts of PERCD SO(8) algebra.

\section{The role of the Foldy--Wouthuysen representation}

Formally there are two different possibilities of introduction of the ERCD and PERCD algebras. These algebras can be put into consideration both in standard PD and FW representation. We put in consideration ERCD algebra (64 orts) and PERCD algebra (29 orts) into the FW representation of the spinor field. Advantages of the FW equation in comparison with the Dirac equation are in coordinate, velocity and spin description [17, 18]. In this representation the equation for the spinor field (the FW equation) has the form

\begin{equation}
\label{eq26} (\partial _0 + i\gamma^{0}\widehat{\omega} )\phi (x)
= 0; \quad \widehat{\omega}\equiv \sqrt{-\Delta +m^{2}}, \; x\in \mathrm{M}(1,3),\; \phi\in \mathrm{H}^{3,4},
\end{equation}

\noindent and is linked with the Dirac equation

\begin{equation}
\label{eq27}(\partial_{0}+iH_{\mathrm{D}})\psi(x)=0,\;H_{\mathrm{D}} \equiv\overrightarrow{\alpha}\cdot\overrightarrow{p}+\beta
m,
\end{equation}

\noindent by the FW transformation $V^{\pm}$:

\begin{equation}
\label{eq28} V^{\pm}\equiv \frac{\pm i\overrightarrow{\gamma}\cdot \nabla+\widehat{\omega}+m}{\sqrt{2\widehat{\omega}(\widehat{\omega}+m)}}.
\end{equation}

\begin{equation}
\label{eq29} \phi(x)=V^{-}\psi(x),\;\psi(x)=V^{+}\phi(x),\;
V^{+}\gamma^{0}\widehat{\omega
}V^{-}=\overrightarrow{\alpha}\cdot\overrightarrow{p}+\beta m.
\end{equation}

Arbitrary operator in PD representation can be found from the operator in FW representation on the basis of the transformation (28): $q^{\mathrm{PD}}=V^{+}q^{\mathrm{FW}}V^{-}$. The important example is the FW spin $\overrightarrow{s}=(s^{mn})=\frac{1}{2}\gamma^{m}\gamma^{n}$. This operator commutes with the FW Hamiltonian $\gamma^{0}\widehat{\omega}$ and in PD representation according to $\overrightarrow{s}^{\mathrm{PD}}=V^{+}\overrightarrow{s}V^{-}$ has the form

\begin{equation}
\label{eq30} \overrightarrow{s}^{\mathrm{PD}}= \overrightarrow{s}-\frac{i[\overrightarrow{\gamma}\times \nabla]}{2\widehat{\omega}} + \frac{\nabla \times [\overrightarrow{s} \times \nabla]}{\widehat{\omega}(\widehat{\omega}+m)}.
\end{equation}

\noindent Spin (30) commutes with the Dirac Hamiltonian $H_{\mathrm{D}}$.

Therefore, in the FW representation the orts of ERCD and PERCD algebras (determining the spin operator) have the simplest form (compare $\overrightarrow{s}=(s^{mn})=\frac{1}{2}\gamma^{m}\gamma^{n}$ with the expression for spin from (30)).

Now the logical necessity of introduction of the ERCD and PERCD algebras in the FW (not PD) representation is evident. The important element of ERCD and PERCD algebras – the spin operator $\overrightarrow{s}=(s^{mn})=\frac{1}{2}\gamma^{m}\gamma^{n}$ in FW representation commutes with the Hamiltonian $\gamma^{0}\widehat{\omega}$ of the FW equation but its commutator with the Dirac Hamiltonian $\overrightarrow{\alpha}\cdot\overrightarrow{p}+\beta
m$ is not equal to zero. However, found by the FW operator $V^{\pm}$ (28) non-local spin (30) commutes with the local Hamiltonian $\overrightarrow{\alpha}\cdot\overrightarrow{p}+\beta
m$ of the Dirac model.

The investigation of FW representation is interest itself in connection with the recent result [19] of V. Neznamov, who developed the formalism of quantum electrodynamics in this representation, see also the results in [20].

Strictly speaking, in (26) and in the theory of the spinor field in general is necessary to appeal to the rigged Hilbert space $\mathrm{S}^{3,4}\subset\mathrm{H}^{3,4}\subset\mathrm{S}^{3,4*}$, where $\mathrm{H}^{3,4}$ is given by (1).  It is necessary due to the fact that the equations (26), (27) have the solutions belonging to the space of the Schwartz generalized functions $\mathrm{S}^{3,4*}$. Therefore, we deal here with the following situation. Strictly speaking, the mathematical correctness of this consideration requires the calculations in the space $\mathrm{S}^{3,4*}$ of the generalized functions, i. e. the application of a cumbersome functional analysis to be made. Nevertheless, let us take into account that the Schwartz test function space $\mathrm{S}^{3,4}$ in the triple $\mathrm{S}^{3,4}\subset\mathrm{H}^{3,4}\subset\mathrm{S}^{3,4*}$ is kernel. This means that $\mathrm{S}^{3,4}$ is dense both in the quantum-mechanical space $\mathrm{H}^{3,4}$ and in the space of generalized functions $\mathrm{S}^{3,4*}$. Therefore, any physical state from $\mathrm{H}^{3,4}$ can be approximated with an arbitrary precision by the corresponding elements of the Cauchy sequence in $\mathrm{S}^{3,4}$, which converges to the given state in $\mathrm{H}^{3,4}$. Further, taking into account the requirement to measure the arbitrary value of the model with non-absolute precision, this means that all necessary calculations can be fulfilled within the Schwartz test function space $\mathrm{S}^{3,4}$ without any loss of generality.

Thus, we use below the Schwartz test function space $\mathrm{S}^{3,4}$ as well as in all our publications on this subject.

\textbf{Maximal pure matrix algebra of invariance of the Foldy--Wouthuysen equation.}

The next important subalgebra of the ERCD algebra is the 32-dimensional algebra

\begin{equation}
\label{eq31} \mathrm{A}_{32}=\mathrm{SO(6)}\oplus i\gamma^{0}\mathrm{SO(6)}\oplus i\gamma^{0},
\end{equation}

\noindent which is the maximal pure matrix algebra of invariance of the FW equation (26). This algebra open the list of subalgebras of 64-dimensional ERCD algebra, which are the symmetries of the Dirac equation in the FW representation. In (31) the generators of SO(6) algebra are known from the formulae (21). (22), (24) and $i\gamma^{0}$ is the Casimir operator of $\mathrm{A}_{32}$ algebra (31).

In PD representation the corresponding algebra $\mathrm{A}_{32}$ found by the help of FW transformation as $q^{\mathrm{PD}}=V^{+}q^{\mathrm{FW}}V^{-}$ still is the symmetry, now of the Dirac equation. However, in the Dirac model the main part of the $\mathrm{A}_{32}$ generators are essentially non-local operators like (30). 

\section{Different representations of the proper extended real Clifford--Dirac algebra}

\textbf{Fundamental Foldy--Wouthuysen representation.}

In the fundamental FW representation the 29 orts of PERCD algebra SO(8) are given by formulae (17), in which the 7 generating $\gamma^{\mathrm{A}}$ operators have the form (15).

\textbf{Standard Pauli--Dirac representation.}

In a standard Pauli--Dirac representation, the so called local representation, the corresponding 29 orts of PERCD algebra SO(8) are the consequences of the FW transformation $V^{\pm}$ (28) and are given by the elements $(\alpha^{\bar{\mathrm{A}}\bar{\mathrm{B}}}=2\widetilde{s}^{\bar{\mathrm{A}}\bar{\mathrm{B}}},
\, \mathrm{I})$, where

\begin{equation}
\label{eq32}
\widetilde{s}^{\bar{\mathrm{A}}\bar{\mathrm{B}}}=\{\widetilde{s}^{\mathrm{A}\mathrm{B}}=\frac{1}{4}[\widetilde{\gamma}
^\mathrm{A},\widetilde{\gamma}
^\mathrm{B}],\,\widetilde{s}^{\mathrm{A}8}=-\widetilde{s}^{8\mathrm{A}}=\frac{1}{2}\widetilde{\gamma}
^\mathrm{A}\}, \quad \bar{\mathrm{A}},\bar{\mathrm{B}}=\overline{1,8},
\, \mathrm{A},\mathrm{B}=\overline{1,7}.
\end{equation}

\noindent Here 7 generating operators $\widetilde{\gamma}
^\mathrm{A}$ together with operators $\widetilde{\gamma}
^\mathrm{0}=V^{+}\gamma^{0}V^{-}$ and
$\widetilde{C}=V^{+}\hat{C}V^{-}$ are nonlocal and have the form

$$\overrightarrow{\widetilde{\gamma}}=\overrightarrow{\gamma}\frac{-\overrightarrow{\gamma}\cdot\nabla+m}{\widehat{\omega}}+\overrightarrow{p}\frac{-\overrightarrow{\gamma}\cdot\nabla+\widehat{\omega}+m}{\widehat{\omega}(\widehat{\omega}+m)},
\,
\widetilde{\gamma}^{4}=\gamma^{4}\frac{-\overrightarrow{\gamma}\cdot\nabla+m}{\widehat{\omega}},$$
\begin{equation}
\label{eq33}
\widetilde{\gamma}^{5}=\widetilde{\gamma}^{1}\widetilde{\gamma}^{3}\widetilde{C},
\,
\widetilde{\gamma}^{6}=i\widetilde{\gamma}^{1}\widetilde{\gamma}^{3}\widetilde{C},
\, \widetilde{\gamma}^{7}=i\widetilde{\gamma}^{0}, \,
\widetilde{\gamma}^{0}=\gamma^{0}\frac{-\overrightarrow{\gamma}\cdot\nabla+m}{\widehat{\omega}},
\end{equation}
$$\widetilde{C}=(\mathrm{I}+2\frac{i\gamma^{1}\partial_{1}+i\gamma^{2}\partial_{2}}{\sqrt{2\widehat{\omega}(\widehat{\omega}+m)}})\hat{C}; \quad \widehat{\omega}\equiv\sqrt{-\triangle+m^{2}}.$$

\textbf{Bosonic representation.}

In bosonic representation, where the proof [2, 21--24] of the
bosonic properties of the FW and Dirac equation is most
convenient, the corresponding 29 orts of the PERCD algebra
(SO(8) algebra) are given by the elements
$(\alpha^{\bar{\mathrm{A}}\bar{\mathrm{B}}}=2\breve{s}^{\bar{\mathrm{A}}\bar{\mathrm{B}}},
\, \mathrm{I})$, where

\begin{equation}
\label{eq34}
\breve{s}^{\bar{\mathrm{A}}\bar{\mathrm{B}}}=\{\breve{s}^{\mathrm{A}\mathrm{B}}=\frac{1}{4}[\breve{\gamma}
^\mathrm{A},\breve{\gamma}
^\mathrm{B}],\,\breve{s}^{\mathrm{A}8}=-\breve{s}^{8\mathrm{A}}=\frac{1}{2}\breve{\gamma}
^\mathrm{A}\},\quad \bar{\mathrm{A}},\bar{\mathrm{B}}=\overline{1,8},
\, \mathrm{A},\mathrm{B}=\overline{1,7}.
\end{equation}

\noindent Here 7 generating operators $\breve{\gamma}
^\mathrm{A}$ have the form

$$\breve{\gamma}^{1}=\frac{1}{\sqrt{2}}\left|
\begin{array}{cccc}
 0 & 0 & 1 & -1\\
 0 & 0 & i & i\\
-1 & i & 0 & 0\\
1 & i & 0 & 0\\
\end{array} \right|,\, \breve{\gamma}^{2}=\frac{1}{\sqrt{2}}\left|
\begin{array}{cccc}
 0 & 0 & -i & i\\
 0 & 0 & -1 & -1\\
-i & 1 & 0 & 0\\
i & 1 & 0 & 0\\
\end{array} \right|,\, \breve{\gamma}^{3} = - \left| {{\begin{array}{*{20}c}
 \sigma ^2 \hfill & 0 \hfill \\
 0 \hfill & {i\sigma ^2} \hfill \\
\end{array} }} \right|\hat{C},$$
\begin{equation}
\label{eq35} \breve{\gamma}^{4} = \left| {{\begin{array}{*{20}c}
 i\sigma ^2 \hfill & 0 \hfill \\
 0 \hfill & {-\sigma ^2} \hfill \\
\end{array} }} \right|\hat{C}, \; \breve{\gamma}^{5}= \frac{1}{\sqrt{2}}\left|
\begin{array}{cccc}
 0 & 0 & -1 & -1\\
 0 & 0 & i & -i\\
1 & i & 0 & 0\\
1 & -i & 0 & 0\\
\end{array} \right|, \; \breve{\gamma}^{6}= \frac{1}{\sqrt{2}}\left|
\begin{array}{cccc}
 0 & 0 & -i & -i\\
 0 & 0 & 1 & -1\\
-i & -1 & 0 & 0\\
-i & 1 & 0 & 0\\
\end{array} \right|, \; \breve{\gamma}^{7}=i\gamma^{0}.
\end{equation}

\noindent It is useful to add to the list (35) the explicit forms in this representation of the following operators:

\begin{equation}
\label{eq36} \breve{\gamma}^{0} = \left| {{\begin{array}{*{20}c}
 \sigma ^3 \hfill & 0 \hfill \\
 0 \hfill & {\sigma ^1} \hfill \\
\end{array} }} \right|, \; \breve{i} = \left| {{\begin{array}{*{20}c}
 i\sigma ^3 \hfill & 0 \hfill \\
 0 \hfill & {-i\sigma ^1} \hfill \\
\end{array} }} \right|, \; \breve{C} = \left| {{\begin{array}{*{20}c}
 \sigma ^3 \hfill & 0 \hfill \\
 0 \hfill & {\mathrm{I}_{2}} \hfill \\
\end{array} }} \right|\hat{C}.
\end{equation}

Transition from the fundamental representation (15), (17) of the
PERCD algebra to the bosonic representation (34)--(36) is fulfilled by the
transformation
$\breve{\gamma}^{\mathrm{A}}=W\gamma^{\mathrm{A}}W^{-1}$ with the
help of the operator $W$:

\begin{equation}
\label{eq37} W = \frac{1}{\sqrt{2}}\left|
\begin{array}{cccc}
 \sqrt{2} & 0 & 0 & 0\\
 0 & 0 & i\sqrt{2}\hat{C} & 0\\
0 & -\hat{C} & 0 & 1\\
0 & -\hat{C} & 0 & -1\\
\end{array} \right|, \; W^{-1}=\left|
\begin{array}{cccc}
 \sqrt{2} & 0 & 0 & 0\\
 0 & 0 & -\hat{C} & -\hat{C}\\
0 & i\sqrt{2}\hat{C} & 0 & 0\\
0 & 0 & 1 & -1\\
\end{array} \right|, \; WW^{-1}=W^{-1}W=\mathrm{I}_{4}.
\end{equation}

\textbf{List of other useful representations.}

The list of different representations, which have been introduced in our investigation of the additional properties of the Dirac and Maxwell equations, contains the representations of PERCD algebra generated by the following $\gamma$  matrices and corresponding transformations.

(i) The representation of $\gamma$  matrices from [9] (formulae (12) in [9]), which has been used for the investigation of the slightly generalized original Maxwell equations and their symmetries.

(ii) The representation of $\gamma$ matrices from [25] (formulae (30) in [25]), which has been used for the construction of the hydrogen atom model on the basis of the slightly generalized original Maxwell equations.

(iii) The transformations from [21] (formulae (21), (30) from [21]), which have been used for the identification of the bosonic solutions of the FW and Dirac equations with nonzero mass.

(iv) The transformation from [2] (formula (15) from [2]), which has been used for the construction of bosonic spin (1,0) representations of the Lorentz and Poincare groups, with respect to which the FW and Dirac equations were proved to be invariant.

(v) The representation of $\gamma$ matrices from [24] (formulae (5) in [24]), which has been used for the axiomatic formulation of the foundations of the relativistic canonical quantum mechanics.

In all these cases (i)--(v) the useful and physically meaningful representations of 29 dimensional PERCD algebra can be found and presented in the completely similar to the given by the formulae (34)--(37) way. Each of representations (i)--(v) of PERCD algebra is related to some quantum mechanical stationary complete set of operators of physical values.

\section{Bosonic spin s=(1,0) Poincar$\mathrm{\acute{e}}$ symmetry of the Foldy-Wouthuysen and Dirac equations}

Below we give the brief illustration of the possibilities and benefit of the ERCD and PERCD algebras. The example of the construction of the bosonic symmetries of the FW and Dirac equation is under consideration. Here we briefly present the main result of our abstract [22].

The fundamental assertion is that subalgebra SO(6) (21)--(23) of PERCD algebra contains two different realizations of SU(2) algebra for the spin s=1/2 doublet. By taking the sum of the two independent sets of SU(2) generators from (21)--(23) one can obtain the SU(2) generators of spin s=(1,0) multiplet, which generate the transformation of invariance of the FW equation (26).

It is easy to show (after our consideration [2]) that $p_{\mu}, \, j_{\mu\nu}$ generators

\begin{equation}
\label{eq38} p_{0} = - i\gamma _{0}\widehat{\omega},\mbox{
}p_{n}=\partial _n,\mbox{ }j_{ln}= x_l
\partial _n -
x_n \partial _l + \breve{s}_{ln}, \, j_{0k}=x_0\partial
_k+i\gamma_{0}\{x_k\widehat{\omega }+\frac{\partial
_k}{2\widehat{\omega }} +\frac{(\overrightarrow{\breve{s}}\times
\overrightarrow{\partial })_k}{\widehat{\omega }+m}\}\\,
\end{equation}

\noindent of group $\mathcal{P}$ commute with the operator of the
FW equation (26) and satisfy the commutation relations of the Lie
algebra of the group $\mathcal{P}$ in manifestly covariant form. Here the spin operators $\overrightarrow{\breve{s}}= \left(\breve{s}_{ln}\right)$ of SU(2), which also commute with the operator of FW equation $\partial _0 + i\gamma^{0}\widehat{\omega}$, are given by

\begin{equation}
\label{eq39} \breve{s}^{1}= \frac{1}{\sqrt{2}}\left|
\begin{array}{cccc}
 0 & 0 & i\hat{C} & 0\\
 0 & 0 & -\hat{C} & 0\\
-i\hat{C} & \hat{C} & 0 & 0\\
0 & 0 & 0 & 0\\
\end{array} \right|, \; \breve{s}^{2}= \frac{1}{\sqrt{2}}\left|
\begin{array}{cccc}
 0 & 0 & \hat{C} & 0\\
 0 & 0 & -i\hat{C} & 0\\
-\hat{C} & i\hat{C} & 0 & 0\\
0 & 0 & 0 & 0\\
\end{array} \right|, \; \breve{s}^{3}= \left|
\begin{array}{cccc}
 -i & 0 & 0 & 0\\
 0 & i & 0 & 0\\
0 & 0 & 0 & 0\\
0 & 0 & 0 & 0\\
\end{array} \right|.
\end{equation}

\noindent The operators (38) generate in the space $\mathrm{H}^{3,4}$ unitary $\mathcal{P}$ representation another than the fermionic $\mathcal{P}^{\mathrm{F}}$-generators D-64 -- D-67 in [18], i. e. the bosonic $\mathcal{P}^{\mathrm{B}}$ representation of the group $\mathcal{P}$, with respect to which the FW equation (26) is invariant. For the generators (38) the Casimir operators have the form:

\begin{equation}
\label{eq40}p^{\mu}p_{\mu}=m^{2}, \,
W^{\mathrm{B}}=w^{\mu}w_{\mu}=m^{2}\overrightarrow{\breve{s}}^{2}=
-1(1+1)m^{2}\left| {{\begin{array}{*{20}c}
 \mathrm{I}_{3} \hfill & 0 \hfill \\
 0 \hfill & {0} \hfill \\
\end{array} }} \right|.
\end{equation}

\noindent Hence, according to the Bargman--Wigner classification,
we consider here the spin s=(1,0) representation of the group
$\mathcal{P}$.

The $\overrightarrow{\breve{s}}= \left(\breve{s}_{ln}\right)$ operators (39) (the generators of the little group for the bosonic Poincar$\mathrm{\acute{e}}$ group $\mathcal{P}$) are the direct consequences of the elements of subalgebra SO(6) (21)--(23) of PERCD algebra. The operators (39) can be presented in the form

\begin{equation}
\label{eq41}\overrightarrow{\breve{s}}\equiv(\breve{s}^{j})=(\breve{s}_{mn})=\frac{1}{2}(\breve{\gamma}^{2}\breve{\gamma}^{3}-\breve{\gamma}^{0}\breve{\gamma}^{2}\breve{C},\,\breve{\gamma}^{3}\breve{\gamma}^{1}+\breve{i}\breve{\gamma}^{0}\breve{\gamma}^{2}\breve{C},\,\breve{\gamma}^{1}\breve{\gamma}^{2}-\breve{i}),
\end{equation}

\noindent where the elements of constructions (41) are given in the formulae (35), (36) of bosonic representation of SO(6) subalgebra of PERCD algebra.

It is important to note that transformation (37) does not change the Hamiltonian of the FW equation (26).

For the Dirac equation (27) the generators of corresponding bosonic $\mathcal{P}^{\mathrm{B}}$ representation of the group $\mathcal{P}$, with respect to which the equation (27) is invariant, can be found easy on the basis of $p_{\mu}, \, j_{\mu\nu}$ generators (38) and FW transformation (28) (see the corresponding procedure in [2]).

Hence, the role of PERCD algebra in the construction of bosonic symmetries of the FW or Dirac equation is shown.

\section{Conclusions}

The 64 dimensional extended real Clifford--Dirac (ERCD) algebra and 29 dimensional proper ERCD (PERCD) algebra, which have been put into consideration in [2], are the useful generalizations of standard 16 dimensional Clifford--Dirac (CD) algebra. Here these algebras are described in details. The different useful representations of such algebras are considered.

Application of ERCD and PERCD algebras enabled us to find [2, 21--23] new bosonic symmetries, solutions and conservations laws for the Dirac equation. These algebras are our main mathematical tool for the demonstration of Fermi--Bose duality of the spinor field and Dirac equation.

The ERCD and PERCD algebras can be used in each model of theoretical physics, where the standard CD was applied. And in each region these algebras open new possibilities in comparison with standard CD algebra.

During the year after the BGL-8 Internationalional Conference on Non-Euclidean Geometry in Modern Physics and Mathematics we used the PERCD algebra in our further proof of the property of the Fermi-Bose duality of the Dirac equation, which was proved on three levels: bosonic symmetries, bosonic solutions and bosonic conservation laws (the corresponding fermionic symmetries, solutions and conservation laws for the Dirac equation are well-known). We also used the PERCD algebra in the construction of relativistic canonical quantum mechanics. These applications of PERCD algebra were recently published in [26--30]. 

The Clifford algebras investigation, description and application are the purpose of special journal Advances in Applied Clifford Algebras, which was organized and developed by Prof. Jaime Keller [31] (unfortunately he passed away on January 7, 2011).

\vskip 1.cm

\end{document}